\documentclass[prl,amsmath,aps,10pt,superscriptaddress,letterpaper,tightenlines, twocolumn]{revtex4}
\usepackage{bm}
\usepackage{stmaryrd}
\usepackage{amsmath}
\usepackage{amssymb}
\usepackage{graphicx}
\usepackage{textcomp}
\usepackage{calrsfs}
\usepackage{yfonts}
\usepackage{amsthm}
\usepackage{lipsum}
\newcommand{\p}{{\partial}}
\newcommand{\pt}{{\partial_t}}

\newcommand{\curl}{{\nabla\times}}
\newcommand{\haf}{{\frac{1}{2}}}
\newcommand{\intf}{{\int_0^{\infty}\,}}
\newcommand{\la}{{\langle}}
\newcommand{\ra}{{\rangle}}

\begin{document}
%%%%%%%%%%%%%%%%%%%%%%%%%%%%%%%%%%%%%%%%%%%%%%%%%%%%%%%%%%%%%%%%%%%
\title{Using quantum friction to synchronize rotating bodies}

\author{Vahid Ameri}
\email{vahameri@gmail.com}
\affiliation{Department of Physics, Faculty of Science, University of Hormozgan, Bandar-Abbas, Iran}

%%%%%%%%%%%%%%%%%%%%%%%%%%%%%%%%%%%%%%%%%%%%%%%%%%%%%%%%%%%%%%%%%%%
\begin{abstract}
Proposing a combined system of a nanoparticle  and a plane surface in the presence of electromagnetic vacuum fluctuations, the electromagnetic and medium fields have been quantized. Quantum friction of nanoparticle due to the presence of plane surface and also electromagnetic vacuum has been discussed. The possibility of synchronizing rotating bodies through the quantum friction between them are investigated. It has been shown that there is a significant connection between the synchronization of the rotating bodies and their quantum friction.
\end{abstract}
%\pacs{12.20.Ds, 42.50.Lc, 03.70.+k}
\maketitle
%%%%%%%%%%%%%%%%%%%%%%%%%%%%%%%%%%%%%%%%%%%%%%%%%%%%%%%%%%%%%%%%%%%
Synchronization as a classical behavior has been observed in a large variety of biological, chemical, physical, and even social context \cite{3sync}. Classical nature of synchronization raised a lot of interests to explore it in the quantum systems \cite{vinokur2008superinsulator,mari2013measures,goychuk2006quantum}. Spontaneous synchronization between two quantum Van der Pol (VdP) oscillators \cite{lee2013quantum,lee2014entanglement,walter2015quantum}, phase locking of a single VdP resonator with an external field \cite{walter2014quantum}, and synchronization between coupled quantum many body systems \cite{qiu2014measure,xu2014synchronization} are the most recent researches on quantum synchronization.
\par Very recently Ameri and et all \cite{ameri2015mutual} introduced the mutual information as an order parameter for quantum synchronization which it can qualify the synchronization on a large variety of systems from semi-classical continuous-variable systems to the deep quantum ones.
\par In this work, we propose a system of a rotating nanoparticle above a semi-infinite dielectric with the planar interface (Figure 1). Using the canonical field quantization approach, we find the explicit form of electromagnetic and medium fields in the nonrelativistic regime, then it is straightforward to derive the quantum friction torque of nano particle due to the presence in the electromagnetic vacuum field and being next to a semi-infinite dielectric bulk .
\par The aim of this work is to discuss the possibility of synchronizing rotating bodies using quantum friction. Quantum friction in the proposed system has a bipartite nature. One part of that corresponds to the friction due to the presence of the electromagnetic vacuum field (self-friction) while the other term is a friction between the nanoparticle and the plane surface. It is straightforward to show that, this term can synchronize the rotation of nanoparticle with the plane surface. Also, it will be shown that there is a significant connection between this quantum friction term and mutual information of the system. 
\par To this aim, we consider the following Lagrangian for a spherical nanoparticle rotating along its symmetric axis (z-axis) with angular velocity $ \omega_0 $, in the vicinity of a semi-infinite dielectric bulk,
\begin{figure}
   \includegraphics[scale=0.5]{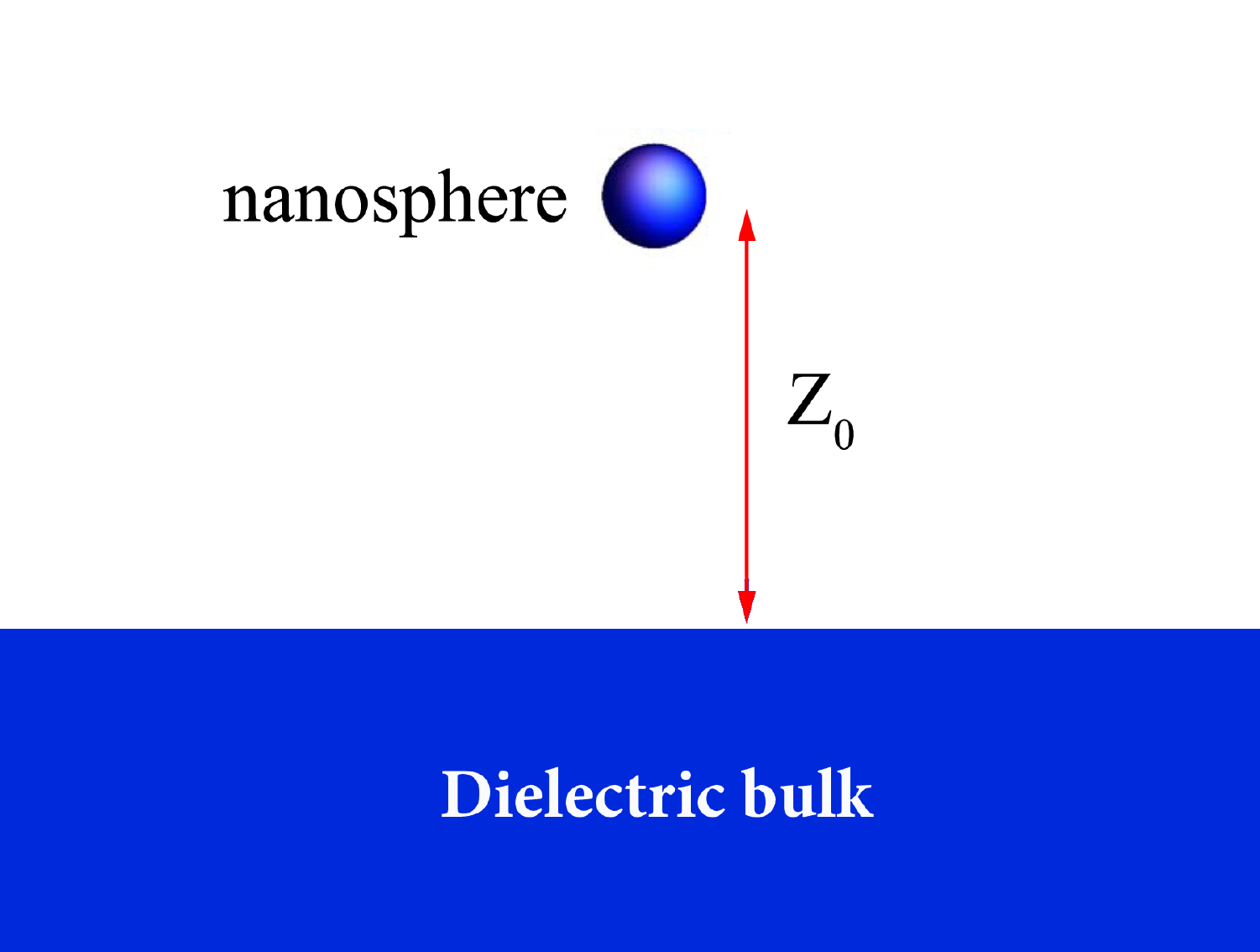}\\
  \caption{A rotating nanoparticle in the vicinity of a semi-infinite homogeneous and isotropic matter }\label{setup}
\end{figure}  
\begin{eqnarray}\label{L}
\mathcal{L} &=& \haf\epsilon_0\,(\pt \mathbf{A})^2-\frac{1}{2\mu_0}(\curl\mathbf{A})^2\nonumber\\
&+&\haf\intf d\nu \,[(\pt \mathbf{X}+\omega_0\p_{\varphi}\mathbf{X})^2-\nu^2\mathbf{X}^2]\nonumber\\
&-& \epsilon_0\intf d\nu\,f_{ij}(\nu,t)X^j\pt A_i\nonumber\\
&+&\epsilon_0\intf d\nu\,f_{ij}(\nu,t)X^j (\mathbf{v}\times\curl\mathbf{A})_i\nonumber\\
&+&\haf\intf d\nu \,[(\pt \mathbf{Y})^2-\nu^2\mathbf{Y}^2]\nonumber\\
&-& \epsilon_0\intf d\nu\,g_{ij}(\nu)Y^j\pt A_i.
\end{eqnarray}
where $ X^j $ and $ Y^j $ are the dielectric fields describing the nanoparticle and the dielectric bulk respectively. the bulk and nanoparticle are considered to be in local thermodynamical equilibrium at temperature $ T $. $ f_{ij}(\nu,t) $ and $ g_{ij}(\nu,t) $ are the correspond coupling tensors between the medium fields and the electromagnetic vacuum field. One can easily derive the response functions of the nanoparticle and the bulk using these coupling tensors \cite{kheirandish2014electromagnetic,ameri2015radiative},
\begin{equation}\label{kapa}
\chi^{0}_{kk}(\omega) =\epsilon_0\intf d\nu\, \frac{f^{2}_{kk} (\nu)}{\nu^2-\omega^2}.
\end{equation}
In nonrelativistic regime, we can ignore the terms containing the velocity and obtain the equations of motion,
\begin{eqnarray}\label{FE}
\mathbf{P}_P (\mathbf{r},\omega)&=&\mathbf{P}_P^{N} (\mathbf{r},\omega)+\epsilon_0 \boldsymbol{\chi}^{P} (\omega,-i\p_{\varphi})\mathbf{E}, \nonumber \\
\Bigl\{\curl\curl &-&\frac{\omega^2}{c^2}\mathbb{I}-\frac{\omega^2}{c^2}\boldsymbol{\chi}^{P}(\omega,-i \p_\varphi) -\frac{\omega^2}{c^2}\boldsymbol{\chi}^{B}(\omega) \Bigl\}\cdot\mathbf{E}\nonumber \\
&& =\mu_0\omega^2(\mathbf{P}_P^N+\mathbf{P}_B^N),
\end{eqnarray}
where $ \mathbf{P}_P^{N} $ and $ \mathbf{P}_B^{N} $ are the fluctuating or noise electric polarization components of the particle and the bulk respectively \cite{kheirandish2014electromagnetic,ameri2015radiative}.
\par Using (\ref{FE}) and the dyadic Green tensor $ G_{ij} $, we find
\begin{eqnarray}\label{e}
E_i (\mathbf{r},\omega) &=& E_{0,i} (\mathbf{r},\omega)+\mu_0\omega^2\int d\mathbf{r}'\,G_{ij} (\mathbf{r},\mathbf{r}',\omega)\,\nonumber \\ && \times (P^{N}_{P,j} (\mathbf{r}',\omega)+P^{N}_{B,j} (\mathbf{r}',\omega)),
\end{eqnarray}
where the first term on the right-hand side of (\ref{e}) corresponds to the fluctuations of the electric field in electromagnetic vacuum, while the second term is the induced electric field due to the presence of the nanoparticle and the bulk.
\par The torque produced by an electric field $ E $ on a rotating particle along its rotation axis $ \hat{z} $ is 
\begin{equation}\label{m}
  \mathbf{M}=\int_V d\mathbf{r} \la \mathbf{p}_p(t)\times \mathbf{E}(\mathbf{r},t) \ra\cdot\mathbf{\hat z}.
\end{equation}
In case of a rotating nanoparticle in the electromagnetic vacuum (absence of plane surface), the frictional torque has been calculated and discussed \cite{kheirandish2014electromagnetic,manjavacas2010vacuum},
 \begin{eqnarray}\label{ms}
    \mathbf{M}_s =\frac{\hbar}{2\pi c^2}\int_0^\infty d\omega \omega^2 \mbox{Im}\,[G_{xx}(\omega) +G_{yy}(\omega) ] \nonumber\\
       \,\bigg[\mbox{Im}\,[\alpha(\omega_+)][a_T(\omega_+)-a_{T_0}(\omega)]\nonumber\\
   -\mbox{Im}\,[\alpha(\omega_-)][a_T(\omega_-)-a_{T_0}(\omega)]\bigg],
 \end{eqnarray} 
 where $ \omega_\pm = \omega \pm \omega_0 $ , $ \mbox{Im}[\alpha(\omega)]= V  \mbox{Im}[\chi(\omega)] $, and $ a_T(\omega)=\coth(\hbar\omega/k_B T) $.
\par The torque (\ref{ms}) generated by the quantum friction of rotating particle due to the presence in electromagnetic vacuum field. In this work, we call it the self-quantum friction. This name does make sense because, in fact, it is a friction between the dipole moment fluctuation of the particle and its induction on the electric field of the vacuum and vice versa.
 While here we are going to derive the quantum friction  between the rotating nanoparticle and the plane surface which it is much bigger than the self-quantum friction.   
 \begin{equation}\label{mb}
  \mathbf{M}_B=\int_V d\mathbf{r} \la \mathbf{p}_p^{ind}(t)\times \mathbf{E}^{ind}(\mathbf{r},t) \ra\cdot\mathbf{\hat z}.
\end{equation}
From (\ref{FE}), $ \mathbf{p}_p^{ind} $ has been written as a function of electric field and the electric field contains the dipole moment fluctuations $ \mathbf{p}_B^{ind} $ of the dielectric bulk. So finally, we derive this torque in term of the dielectric bulk dipole moment fluctuation $\la P^N_{B,i}(\mathbf{r},\omega)\cdot P^N_{B,j}(\mathbf{r'},\omega')\ra$.
One can say, it is the quantum friction torque of nanoparticle due to presence next to a plane surface. 
\par The friction between two bodies has a bipartite nature, on the other hand, while this friction is going to stop the moving body, it forces the other body to make a rotation. This interesting future of the friction possibly can make it a good candidate to synchronize the rotating bodies. Although the quantum friction is a very small effect, the synchronization phenomena need such a small force to happen. In the following, we try to find more evidence on the ability of quantum friction on synchronizing the rotating bodies.
\par Using the fluctuation-dissipation relation,
 \begin{eqnarray}
 \la P^{N}_{Bi}(\mathbf{r},\omega)P^{N\dag}_{Bj}(\mathbf{r}',\omega')\ra = 8\pi\epsilon_0 \hbar\, \mbox{Im}[\chi^{B}_{ij} (\omega)]\nonumber \\
 \times  n_T (\omega) \,\delta(\mathbf{r}-\mathbf{r}')\delta(\omega-\omega'),
 \end{eqnarray}
and (\ref{mb}), we find the quantum friction torque between the plane surface and nanoparticle
\begin{eqnarray}
\mathbf{M}_B=\int d\omega \int d\mathbf{r'}\frac{\hbar \pi^2 \omega}{c^4}\mbox{Im}[\alpha^p(\omega_+)-\alpha^p(\omega_-)]\nonumber \\
 \mbox{Im}[\chi^B(\omega)]n_T(\omega) \{G_{yi}(\mathbf{r},\mathbf{r'},\omega)G_{yi}^*(\mathbf{r},\mathbf{r'},\omega) \nonumber \\
 +G_{xi}(\mathbf{r},\mathbf{r'},\omega)G_{xi}^*(\mathbf{r},\mathbf{r'},\omega)\},
\end{eqnarray}
where $ n_T(\omega)=1/(e^{-\beta \hbar \omega}-1) $.
\par To have some numerical results, the dyadic Green tensor for this geometry\cite{maradudin1975scattering} has been substituted and both dielectric bulk and nanoparticle are considered to be made of Silicon Carbide (SiC) where the dielectric function is given by the oscillator model \cite{palik1998handbook},
\begin{equation}
\varepsilon(\omega)=\varepsilon_\infty(1+\frac{\omega_L^2-\omega_T^2 }{\omega_T^2-\omega^2-i\Gamma \omega }),
\end{equation}
with $\varepsilon_\infty=6.7$, $\omega_L=1.823\times 10^{14}$, $\omega_T=1.492\times 10^{14}$, and $\Gamma = 8.954\times 10^{11}$.
\begin{figure}
   \includegraphics[scale=0.6]{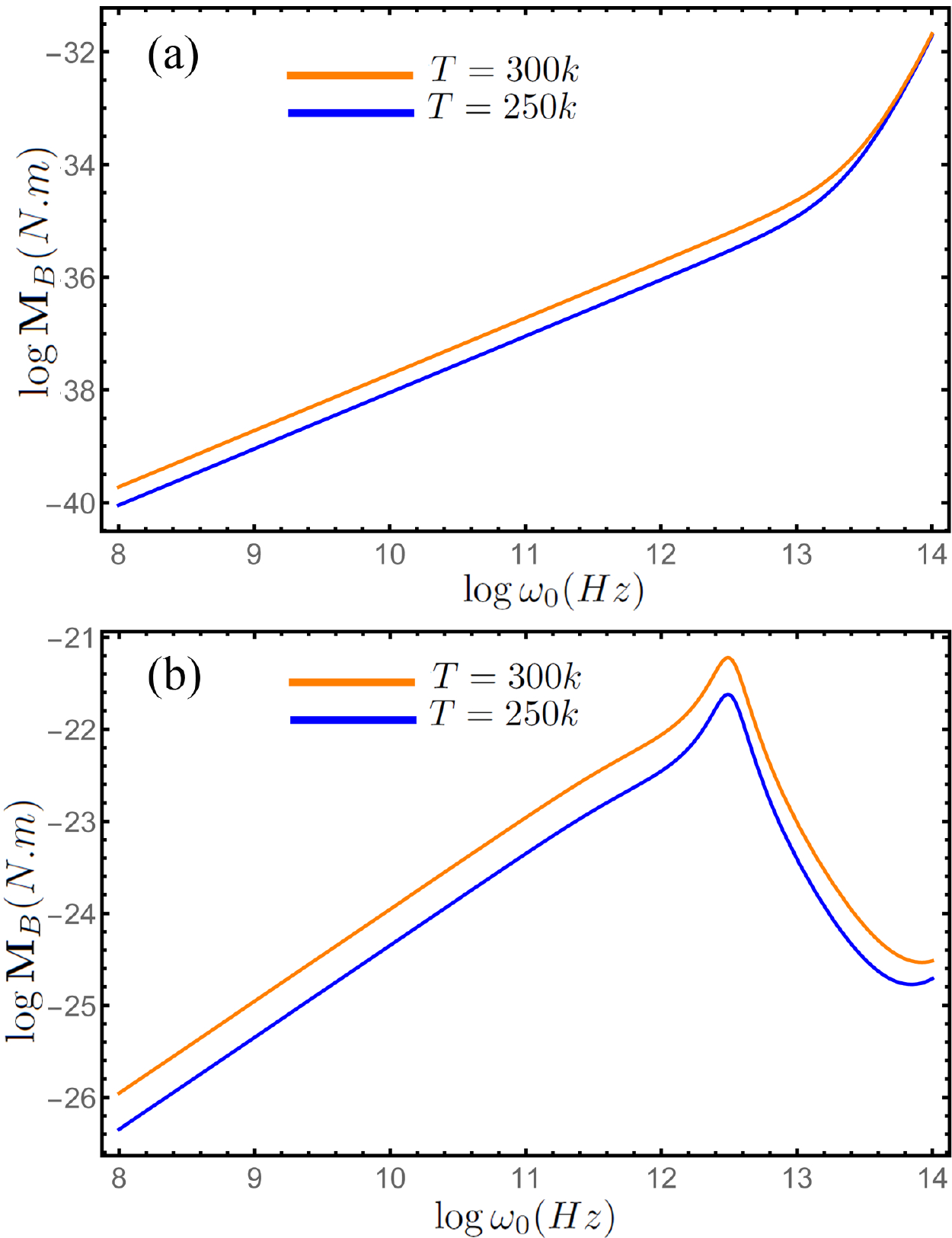}\\
  \caption{(Color online) Logarithmic (base 10) plot of the quantum friction torque of a rotating nanoparticle in (a) electromagnetic vacuum field (self-quantum friction), and (b) the presence of a plane surface, as a function of the angular velocity $ \omega_0 $ of the nanoparticle for two temperatures ($ T=250 k $ and $ T=300 k $)  }\label{mbw0}
\end{figure}
\par In figure (\ref{mbw0}), the quantum frictional torque of a single rotating nanoparticle and a rotating nanoparticle in the presence of a plane surface has been depicted as a function of the angular velocity of rotating nanoparticle $ \omega_0 $ for different temperatures. As mentioned before, the self-quantum friction torque is much smaller than the frictional torque in the presence of semi-infinite dielectric bulk. Also, one can see another interesting future, comparing the quantum friction torques in figure (\ref{mbw0}a) and figure (\ref{mbw0}b), while for $ \omega_0\ll \Gamma $ they both behave as similar functions of angular velocity of nanoparticle, the frictional torque $ \mathbf{M}_B $ (in the presence of dielectric bulk) has a remarkable peak for $ \omega_0\approx 3\times 10^{12} $ where it varies with the dielectric function and also very slowly with the temperature.
\begin{figure}
   \includegraphics[scale=0.4]{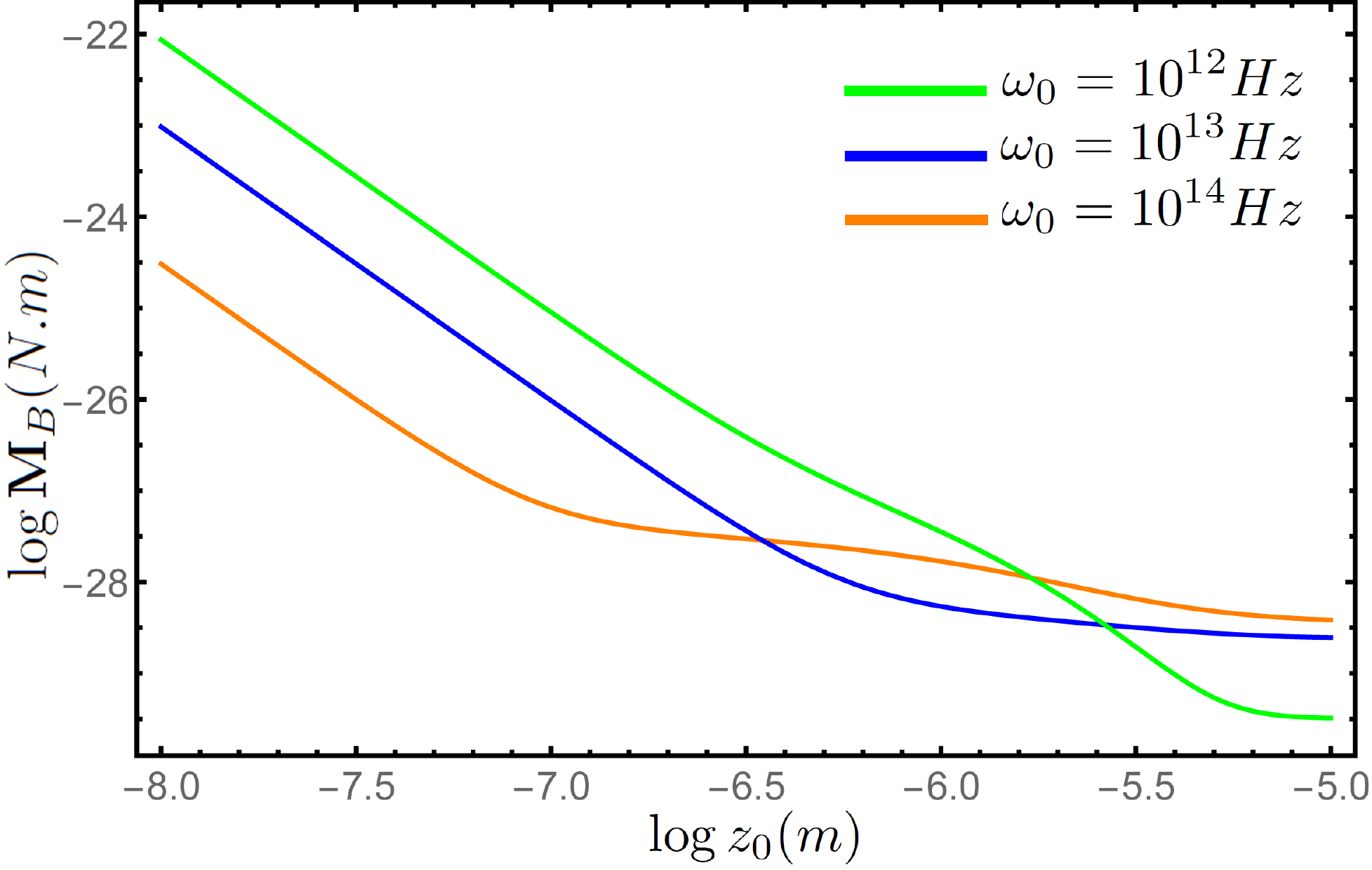}\\
  \caption{(Color online) Logarithmic (base 10) plot of the quantum friction torque of a rotating nanoparticle in the presence of a plane surface, as a function of the distance $ z_0 $ to the surface for different angular velocities $ \omega_0 $ of the nanoparticle. }\label{mbz}
\end{figure} 
\par Figure (\ref{mbz}) shows the frictional torque  $ \mathbf{M}_B $ of the rotating nanoparticle in term of the distance $ z_0 $ to the plane surface. The frictional torque increases as $ 1/z_0^3 $ at small distances, less than $ 100 nm $, from the surface known as near field effect \cite{betzig1992near}. One can propose that, to have more chances on synchronizing the rotation of the bodies, it is better to be at small distances from the surface called near field zone.
\par There is an interesting report on the entropy of a system with a nanoparticle and plane surface \cite{pendry1999radiative,pendry1983quantum} where there were shown that the maximum flow of entropy in a single channel is linked to the flow of energy. Also separately in \cite{ameri2015radiative}, it has been discussed in details, the heat flow between a rotating nanoparticle and plane surface has been reported to increase as  $ 1/z_0^3 $ in the near field zone.
\par Mutual information of system $AB$ composed of two subsystem $A$ and $B$ is defined as
\begin{equation}
I=S_A+S_B-S_{AB}
\end{equation} 
where $ S_A $ and $ S_B $ are the entropies of subsystem $ A $ and $ B $ respectively and $ S_{AB} $ shows the total entropy of the system. Mutual information qualifies how much the knowledge of the subsystem $ A $ gives information about the subsystem $ B $ \cite{cover1991entropy}. So, increasing the flow of entropy between the subsystems increases the mutual information of the system.
\par As mentioned before, mutual information has been introduced as an order parameter for quantum synchronization \cite{ameri2015mutual}, where increasing in the mutual information considered as a signal of the presence of the quantum synchronization. Interestingly here, getting closer to the surface, the quantum friction, channel of making rotational synchronization, increases and spontaneously increasing of the mutual information signaling the presence of a synchronization.
\par finally to be more clear about the synchronization in the introduced system, let's say, the real synchronization would happen between a rotating micro particle with angular velocity $ \omega_0 $ and a nanoparticle, rotating above it by the angular velocity $ \omega_0+\delta $. The case is so similar to our considerations because a micro particle could be considered as an infinite plane surface for a nanoparticle. The major difference between the systems is the rotation of the plane surface, while for simplicity we consider it to be static. But, the point is, fundamental aspect of quantum friction torque will remain unchanged because there is a same relative motion between the nanoparticle and the plane surface in both cases.
\bibliography{sci}
\bibliographystyle{apsrev4-1}
\end{document}